\documentclass{article}
\usepackage{amsmath, amsfonts, amsrefs}
\usepackage{url} 

\newcommand{\bbC}{{\mathbb{C}}}
\newcommand{\cE}{{\mathcal{E}}}

\newcommand{\br}{{\mathbf{r}}}

\newcommand{\tr}{\mathop{\mathrm{tr}}}

\newcommand{\poly}{\mathop{\mathrm{poly}}}
\DeclareMathOperator{\triv}{trivial}

\def\Uqft{U_{\text{QFT}}}
\def\Ucg{U_{\text{CG}}}

\renewcommand{\>}{\rangle}
\newcommand{\<}{\langle}
\newcommand{\ket}[1]{|#1\rangle}
\newcommand{\bra}[1]{\langle #1|}

\newcommand{\proj}[1]{\left|#1\right\>\!\left\<#1\right|}
\newcommand{\ot}{\otimes}

\def\benum{\begin{enumerate}}
\def\eenum{\end{enumerate}}
\def\bitem{\begin{itemize}}
\def\eitem{\end{itemize}}

\newcommand{\be}{\begin{equation}}
\newcommand{\ee}{\end{equation}}
\def\ba#1\ea{\begin{align}#1\end{align}}

\newcommand{\eq}[1]{(\ref{eq:#1})}

\begin{document}
\title{Quantum expanders from any classical Cayley graph expander}
\author{Aram W. Harrow\\
Department of Computer Science \\
University of Bristol \\
Bristol, UK\\
a.harrow@bris.ac.uk}
\date{\today}
\maketitle
\begin{abstract}
  We give a simple recipe for translating walks on Cayley graphs of a
  group $G$ into a quantum operation on any irrep of $G$.  Most
  properties of the classical walk carry over to the quantum
  operation: degree becomes the number of Kraus operators, the
  spectral gap becomes the gap of the quantum operation (viewed as a
  linear map on density matrices), and the quantum operation is efficient
  whenever the classical walk and the quantum Fourier transform on $G$
  are efficient.  This means that using classical constant-degree
  constant-gap families of Cayley expander graphs on e.g. the
  symmetric group, we can construct efficient families of quantum
  expanders.
\end{abstract}

{\bf Background.}  Classical expanders can be defined in either
combinatorial or spectral terms, while quantum expanders usually have
only a spectral definition.  Quantum expanders were introduced in
\cite{Hastings-ent-expand} for their application to quantum spin
chains and in \cite{BT-expand-QSZK} for applications to quantum
statistical zero knowledge.  Here we (following
\cite{Hastings-ent-expand} and \cite{BST-explicit}) define a $(N, D, \lambda)$
quantum expander to be a quantum operation $\cE$ that 
\bitem
\item Has $N$-dimensional input and output.
\item Has $\leq D$ Kraus operators.
\item Has second-largest singular value $\leq \lambda$.  Equivalently,
  if $\cE(\rho)=\rho$ and $\tr\rho\sigma=0$ then 
  $\|\cE(\sigma)\|_2 \leq \lambda\|\sigma\|_2$, where $\|X\|_2 :=
  \sqrt{\tr X^\dag X}$. 
\eitem
We say that $N$ is the dimension of the expander, $D$ its degree (by
analogy with classical expanders) and $1-\lambda$ its gap.  Note that
all quantum operations have at least one fixed state and thus at least
one eigenvalue equal to one.  The above definition is stricter than the one
in \cite{BT-expand-QSZK}, which demanded only that an expander
increase the von Neumann entropy of a state by at most a constant
amount.  Finally, we say that an expander is efficient (or
``explicit'') if it can be implemented on a quantum computer in time
$\poly(\log N)$.  This paper will describe a new method for
constructing quantum expanders, which will in some cases yield
efficient $(N, O(1), \Omega(1))$ expanders for all values of $N>1$.

{\bf Previous work on efficient quantum expanders.} 
  In \cite{Hastings-RU-expand} it was shown that, just
as random constant-degree graphs are likely to be expander graphs,
quantum operations that apply one of a constant number of random
unitaries are likely to be quantum expanders, with nearly the optimal
spectral gap for any fixed degree.  Naturally such expanders cannot be
efficiently constructed; in fact, the best deterministic construction for
them\cite{BST-explicit} takes time exponential in the dimension $N$,
which is doubly-exponential in the number of qubits.

Prescriptions for potentially efficient constructions are given in
\cite{Hastings-ent-expand} and \cite{BT-expand-QSZK}.  Both begin with
classical expanders and turn them into quantum expanders.  The
proposal in \cite{Hastings-ent-expand} is to start with a so-called
``tensor power expander'' and then to add phases.  A tensor product
expander is a degree $D$ graph $(V,E)$ where: (a) each outgoing edge
is labelled $1,\ldots,D$, and (b) if $G'$ is the graph with vertices $V\times V$
and edges given by all pairs $(e_1,e_2)\in E\times E$ such that $e_1$
and $e_2$ have the same label, then $G'$ is an expander.
  Unfortunately, when Cayley graphs are
labeled in the natural way (with label $g$ corresponding to
multiplication by group element $g$) they are not tensor power
expanders.  It seems plausible that random constant-degree graphs
would be tensor power expanders, but this has not been proven.

The approach of \cite{BT-expand-QSZK} is, like this paper, to turn
classical Cayley graph expanders into quantum expanders.  Its main
idea is to apply a classical expander twice: first in the standard
basis, and then conjugated by a sort of generalized Hadamard transform
(which they call a ``good basis change''), so that it acts in a
conjugate basis.  Unfortunately, the quantum Fourier transform is not,
by itself, always enough to make a good basis change.  For some
groups, such as $SL(2,q)$, it is, and thus \cite{BT-expand-QSZK}
obtain a quantum expander based on the classical LPS expander graph.
However, it is unknown how to perform the QFT on $SL(2,q)$ efficiently
(see \cite{MRR04} for partial progress), and so we do not know how to
efficiently perform the basis change required for their construction.
On the other hand, while there are groups such as $S_n$ for which both
efficient QFT's and explicit constant-degree expanders are known, none
have yet been proved to satisfy the additional property needed for the
QFT to be a good basis change.

Very recently, two different constructions of efficient,
constant-degree quantum expanders have appeared.  The first is
described 
in\cite{BST-explicit}.  Their approach is to
generalize the classical zig-zag product\cite{RVW-zig-zag} to quantum
expanders, using a constant number of random
unitaries\cite{Hastings-RU-expand} for the base case.  Like our paper,
\cite{BST-explicit} also describes a family of constant-degree,
constant-gap, efficient expanders.  A minor advantage of our
construction is that it can be made to work for any dimension $N>1$,
while \cite{BST-explicit} requires that $N$ be of the form $D^{8t}$
for a positive integer $t$ and that $D>D_0$ for a universal
constant  $D_0$.

Another efficient constant-degree expander is given in
\cite{margulis-wigner}.  Their approach is to turn the classical
Margulis expander\cite{margulis} into an operation on quantum phase
space.  This results in quantum expanders with the same parameters as
the Margulis expander (degree 8, second largest eigenvalue
$\lambda\leq 2\sqrt{5}/8$) in any dimension, including even infinite
dimensional systems.  While their paper only describes an efficient
construction for dimensions of the form $N=d^n$ for small $d$, their
approach is easily generalized to run in time $\poly\log N$ for any
$N$.

Finally, if we relax the assumption that expanders have constant
degree, then efficient constructions have been described in
\cites{AS04,DN06}.

{\bf Representation theory notation.}  Let $G$ be a group (either
finite or a compact Lie group), and $\hat{G}$ a complete set of
inequivalent unitary irreducible representations (irreps).  For an
irrep $\lambda\in\hat{G}$ and a group element $g \in G$, we denote the
representation matrix by $\br_\lambda(g)$, its dimension by
$d_\lambda$ and the space it acts upon by $V_\lambda$.  Let $\Uqft$ be
the Fourier transform on $G$, corresponding to the isomorphism
$$\bbC[G] \cong \bigoplus_\lambda V_\lambda \ot V_\lambda^*.$$
It is given by the explicit formula $\Uqft = \sum_{g,\lambda,i,j}
\sqrt{d_\lambda/|G|}\br_\lambda(g)_{i,j} \ket{\lambda,i,j}\bra{g}$.
Let $L_x:=\sum_{g\in G}\ket{xg}\bra{g}$ denote the left
multiplication operator.  Then in the Fourier basis, this translates
into action on the first tensor factor.
\be \Uqft L_x \Uqft^\dag = \sum_{\lambda\in\hat{G}}
\proj{\lambda} \ot \br_\lambda(x) \ot I_{d_\lambda}.
\label{eq:FT-left-reg-rep}\ee

{\bf Expander construction.}
Let $\Gamma\subset G$ be the set of generators for a Cayley graph on
$G$.  Choose any non-trivial $\lambda\in\hat{G}$.  Define a quantum
operation $\cE$ on $V_\lambda$ by
\be\cE(\rho) = \frac{1}{|\Gamma|} \sum_{g\in \Gamma} \br_\lambda(g)
\rho \br_\lambda(g)^\dag.\label{eq:my-q-expander}\ee
I claim that
\benum
\item The degree of $\cE$ is $\leq |\Gamma|$. 
\item If (a) group multiplication in $G$ is efficient, (b) there is a
  procedure for efficiently sampling from $\Gamma$, (c) the QFT on $G$
  is efficient and (d) $\log |G|\leq \poly(\log d_\lambda)$, then $\cE$ can be
  implemented efficiently.  
\item
\be \lambda_2(\cE) \leq \lambda_2(W_\Gamma).\label{eq:gap-ineq}\ee
  Here $\lambda_2(\cE)$
  is the second largest singular value of $\cE$, when interpreted as a
  linear map on density matrices, while $\lambda_2(W_\Gamma)$ is the
  second-largest singular value of the Cayley graph transition matrix:
$$W_\Gamma = \frac{1}{|\Gamma|} \sum_{\gamma\in\Gamma}\sum_{g\in G}
\ket{\gamma g}\bra{g}.$$
\eenum
Thus, classical Cayley graph expanders give quantum expanders. 

{\bf Proof of spectral gap.} The first claim is immediate.  In the
second claim, condition (d) is because we say the QFT on $G$ is
efficient if it runs in time $\poly(\log |G|)$, but we would like our
expander to run in time $\poly(\log d_\lambda)$.  Alternatively (a),
(c) and (d) can be replaced by any other efficient procedure for
performing $\br_\lambda(g)$ on a quantum computer.

The only non-trivial claim above is 
\eq{gap-ineq}.  First observe that the stationary state of $W_\Gamma$ is
the uniform distribution
$$\ket{u}:=\frac{1}{\sqrt{|G|}} \sum_{g\in G}\ket{g}.$$
We can find the second largest eigenvalue by subtracting off
a projector onto the stationary state and taking the operator norm.
Thus
\be\lambda_2(W_\Gamma) = \| W_\Gamma - \proj{u} \|_\infty,
\label{eq:cl-gap-def}\ee
where $\|M\|_\infty$ is the largest singular value of $M$.

Similarly, the maximally mixed state $\tau:=I_{d_\lambda} /\sqrt{d_\lambda}$
is a stationary state of $\cE$.  We choose the normalization so that
$\tau$ will be a unit vector with respect to the 
Hilbert-Schmidt 
inner product $\<A,B\> := \tr A^\dag B$.
However, to analyze $\cE$ as a
linear operator, it is simpler to think of it as acting on
vectors.  The corresponding linear map is denoted $\hat{\cE}$ and is
defined to be
\be \hat{\cE} := \frac{1}{|\Gamma|}
\sum_{\gamma\in\Gamma} \br_\lambda(\gamma) \ot
\br_\lambda(\gamma)^*,\ee
where the $^*$ denotes the entry-wise complex conjugate with respect to a
basis $B_\lambda$ for $V_\lambda$.  Then 
$\ket{\hat{\tau}}:=d_\lambda^{-1/2}\sum_{b\in B_\lambda}\ket{b}\ot\ket{b}$ is a
fixed point of $\hat{\cE}$.  Thus
\be \lambda_2(\cE) = \| \hat{\cE} - \proj{\hat{\tau}}\|_\infty.
\label{eq:q-gap-def}\ee

We now use representation theory to analyze \eq{cl-gap-def} and
\eq{q-gap-def}.  First, examine \eq{cl-gap-def}.  Since $\Uqft$ is
unitary, $\| W_\Gamma - \proj{u} \|_\infty = \| \Uqft W_\Gamma\Uqft^\dag -
\Uqft\proj{u}\Uqft^\dag\|_\infty$.  Since $\Uqft\ket{u}=\ket{\triv}$, we can use
\eq{FT-left-reg-rep} to obtain
\ba \lambda_2(W_\Gamma) &= \| W_\Gamma - \proj{u} \|_\infty
= \left\| \frac{1}{|\Gamma|}\sum_{\gamma\in\Gamma}
\sum_{\lambda\in\hat{G}} \proj{\lambda} \ot \br_\lambda(\gamma) \ot
I_{d_\lambda} - \proj{\triv}\right\|_\infty \\
& = 
\left\| \frac{1}{|\Gamma|}\sum_{\gamma\in\Gamma}
\sum_{\substack{\lambda\in\hat{G}\\\lambda\neq\triv}} 
\proj{\lambda} \ot \br_\lambda(\gamma) \ot
I_{d_\lambda} \right\|_\infty \\
& = \max_{\lambda\neq\triv}
\left\| \frac{1}{|\Gamma|}\sum_{\gamma\in\Gamma}
\br_\lambda(\gamma)\right\|_\infty
\label{eq:cl-lambda2-irrep}\ea

A similar argument applies to \eq{q-gap-def} as well.  Here the first
step is to decompose $V_\lambda\ot V_\lambda^*$ into irreps of $G$.
In general,
$$V_\lambda \ot V_\lambda^* \cong \bigoplus_{\nu\in\hat{G}} V_\nu \ot
\bbC^{m_\nu},$$ where $m_\nu$ is the multiplicity (possibly zero) of $V_\nu$ in 
$V_\lambda \ot V_\lambda^*$.  Let $\Ucg$ be the unitary transform
implementing the above isomorphism.  Then by definition,
\be \Ucg \left(\br_\lambda(g) \ot \br_\lambda(g)^*\right)
\Ucg^\dag = \sum_{\nu\in\hat{G}} \proj{\nu} \ot \br_\nu(g) \ot
I_{m_\nu}.\ee
We can use this to analyze the spectrum of $\cE$.  In particular
\be \Ucg \hat{\cE} \Ucg^\dag = \sum_{\nu\in\hat{G}} \proj{\nu}
\ot \left(\frac{1}{|\Gamma|} \sum_{\gamma\in\Gamma}
\br_\nu(\gamma)\right) \ot I_{m_\nu}.\ee
From Schur's Lemma, we know that $m_{\triv}=1$, corresponding to the
stationary state $\ket{\hat{\tau}}$.  Thus
\ba \lambda_2(\cE) &= \|\cE - \proj{\hat{\tau}}\|_\infty \\
& = \|\Ucg(\cE - \proj{\hat{\tau}})\Ucg^\dag\|_\infty \\
& = \max_{\substack{m_\nu\neq 0\\\nu\neq\triv}} 
\left\|\frac{1}{|\Gamma|} \sum_{\gamma\in\Gamma}\br_\nu(\gamma)
\right\|_\infty \\
& \leq \max_{\substack{\nu\neq\triv}} 
\left\|\frac{1}{|\Gamma|} \sum_{\gamma\in\Gamma}\br_\nu(\gamma)
\right\|_\infty\\
& = \lambda_2(W_\Gamma).\ea
This completes the proof.

{\bf Examples of quantum expanders.}  If $G=S_n$ then we can use the
explicit expander of \cite{Kass} and the efficient QFT of
\cite{Bea97}.  The dimension $N=d_\lambda$ can be the size of any irrep of
$S_n$, which asymptotically can be as large as
$\sqrt{n!}\exp(-O(\sqrt{n}))$.  Run-time is thus poly-logarithmic in
the dimension, meaning polynomial in the number of qubits.  However if
we would like an expander on exactly $N$ dimensions, we are not
guaranteed that $n\leq \poly\log(N)$ exists such that $d_\lambda=N$
for some $\lambda\in\hat{S}_n$, nor do we know how to efficiently
check, for a given $n$, whether such a $\lambda$ exists.  (For
completeness, we mention here that irreps of $S_n$ are labeled by
partitions $(\lambda_1,\ldots,\lambda_n)$ with
$\lambda_1+\ldots+\lambda_n=n$ and $\lambda_1\geq \ldots \geq
\lambda_n \geq 0$.  Their dimension is given by $d_\lambda = n!
\prod_{i<j} (\lambda_i - \lambda_j -i + j) / \prod_i (\lambda_i + n -
i)!$.) 

Some other Cayley graph constructions also carry over.  For example,
the (classical) zig-zag product can be interpreted as a Cayley graph,
where the group is an iterated wreath product\cite{RSW-cayley}.
Additionally, the irreps of these wreath products are large (although
also with possibly inconvenient dimensions) and quantum Fourier
transforms on them can be performed efficiently\cite{MRR04}.  Thus,
classical zig-zag product expanders can also be used to construct
efficient, constant-degree, constant-gap quantum expanders.  (We
remark in passing that this construction appears not to be related to
the quantum zig-zag product of \cite{BST-explicit}.)

If we permit approximate constructions then we can relax the
assumption that $G$ is finite.  For example, if $G=SU(2)$ then several
explicit expanders are known\cites{GS-expander, GJS-SU2}, but no
efficient circuits are yet known for the 
QFT.  It would suffice even to be able to implement $\br_\lambda(g)$
in time poly-logarithimic in $d_\lambda$.  This latter result is
claimed by \cite{Zalka04}, but the algorithm 
there is missing crucial steps.

Finally, to construct expanders for any dimension $N>1$ we can use
the fact that the $S_{N+1}$-irrep $\lambda=(N,1)$ has dimension $N$.
To implement $\br_\lambda(\pi)$ for $\pi\in S_{N+1}$ we cannot use the
QFT on $S_{N+1}$, since our run-time needs to be $\poly\log(N)$.
However, we can instead embed $V_\lambda$ into the $N+1$-dimensional
defining representation of $S_{N+1}$, which is given by
$\br_{\text{def}}(\pi)\ket{x} = \ket{\pi(x)}$ for $x=1,\ldots,N+1$.  This
representation is reducible and decomposes into one copy of trivial
representation (spanned by $\ket{1}+\ldots+\ket{N+1}$) and one copy of
the $N$-dimensional irrep $V_{(N,1)}$.  To embed $V_\lambda$ in the
defining representation, we can use any $N+1$-dimensional unitary that
maps $\ket{N+1}$ to $\frac{1}{\sqrt{N+1}}\sum_{x=1}^{N+1}\ket{x}$.
Then performing $\br_{\text{def}}(\pi_j)$ (for Cayley graph generator $\pi_j$)
requires only that $\pi_j(x)$ be computable from $j$ and $x$ in time
$\poly(\log N)$.  A careful examination of the construction of
\cite{Kass} shows this to be the case.  Thus, this technique yields
constant-degree, constant-gap explicit expanders for any dimension
$N>1$.  (Of course, for low enough values of $N$ the degree will be
larger than $N^2$ and so the resulting expander will be inferior to
the trivial ``expander'' which applies a random generalized Pauli
matrix.)

{\bf Acknowledgments.}  I would like to thank Avi Ben-Aroya for useful
comments on the first arXiv version of this paper, Matt Hastings for many
interesting conversations on this subject, and Cris Moore for
pointing out \cite{Kass} and crucially asking why any classical
expander couldn't be turned into a quantum expander.  I am also
grateful to the Oza family for their kind hospitality while I did most
of this work.  My funding is from the Army Research Office under grant
W9111NF-05-1-0294, the European Commission under Marie Curie grants
ASTQIT (FP6-022194) and QAP (IST-2005-15848), and the U.K. Engineering
and Physical Science Research Council through ``QIP IRC.''

\end{document}